\documentclass[times, review, 10pt]{elsarticle}
\usepackage{geometry}
\usepackage{amssymb}
\usepackage{subfigure}
\usepackage{longtable}
\usepackage{multirow}
\usepackage{mathrsfs}
\usepackage{amsfonts}
\usepackage{amsmath}
\usepackage{array}
\usepackage{color}
\usepackage[ruled]{algorithm2e}
\usepackage{float}
\biboptions{numbers,sort&compress}
\usepackage{hyperref}

\begin{document}

\begin{frontmatter}



\title{Audios Don't Lie: Multi-Frequency Channel Attention Mechanism for Audio Deepfake Detection}


\author{Yangguang Feng}
\address{School of Information Engineering, Zhongnan University of Economics and Law, Wuhan, 430073, China}
\ead{fengyg@stu.zuel.edu.cn}

\begin{abstract}
	
	With the rapid development of artificial intelligence technology, the application of deepfake technology in the audio field has gradually increased, resulting in a wide range of security risks. Especially in the financial and social security fields, the misuse of deepfake audios has raised serious concerns. To address this challenge, this study proposes an audio deepfake detection method based on multi-frequency channel attention mechanism (MFCA) and 2D discrete cosine transform (DCT). By processing the audio signal into a mel spectrogram, using MobileNet V2 to extract deep features, and combining it with the MFCA module to weight different frequency channels in the audio signal, this method can effectively capture the fine-grained frequency domain features in the audio signal and enhance the Classification capability of fake audios. Experimental results show that compared with traditional methods, the model proposed in this study shows significant advantages in accuracy, precision,recall, F1 score and other indicators. Especially in complex audio scenarios, this method shows stronger robustness and generalization capabilities and provides a new idea for audio deepfake detection and has important practical application value. In the future, more advanced audio detection technologies and optimization strategies will be explored to further improve the accuracy and generalization capabilities of audio deepfake detection.
	
\end{abstract}


\end{frontmatter}


\section{Introduction}

With the rapid development of artificial intelligence technology, deepfake has become a widely-watched synthetic media technology. Deepfake technology can generate highly realistic fake images, videos, and audios through advanced machine learning algorithms \cite{10323004}such as generative adversarial networks (GANs)\cite{10226172}, autoencoders, and diffusion models\cite{10678647}. These synthetic media are widely used in the entertainment industry, such as filmmaking, video games, and other fields, providing creators with a wealth of creative means. However, as this technology matures, deepfake is gradually being used for malicious purposes, posing great security risks to society.

Deepfake audios, especially their abuse in the financial and social security fields, have attracted widespread attention. The most typical case is that on February 4, according to the Hong Kong Wen Wei Po, a fraud case that shocked Hong Kong revealed the dark side of deepfake technology. The fraud group used AI "deepfake" technology\cite{10447923} to defraud a Hong Kong branch of a multinational company and successfully defrauded a huge amount of money up to 200 million Hong Kong dollars. Such incidents not only expose the potential risks of audio deepfake technology, but also make people aware of its terrible application in crimes such as identity theft and fraud. Especially today when information dissemination and public trust are facing severe challenges, deepfake technology poses an unprecedented threat to media security, political opinion and social stability.

To meet this challenge, researchers have begun to focus on deepfake detection technology. Most of the existing deepfake detection methods focus on the detection of image and video forgery\cite{10718393}, while deepfake of audios lacks efficient detection methods due to their particularity. Audio deepfake detection often rely on deep neural networks to model audio features, but traditional methods cannot effectively capture the complex temporal dependencies\cite{wan2022robust} and frequency characteristics in audio signals\cite{croitoru2024deepfakemediagenerationdetection}. Therefore, how to accurately identify true and false audios has become one of the difficulties in current technical research.

Therefore,this study proposes an audio deepfake detection model based on multi-frequency channel attention mechanism  and 2D discrete cosine transform. The contributions of this study are mainly reflected in the following aspects: (1) A new audio deepfake detection method combining 2D discrete cosine transform and multi-frequency channel attention mechanism is proposed, which can effectively capture the fine-grained features in the audio frequency domain, further improving the robustness and accuracy of audio deepfake detection; (2) Through the innovative design of MobileNet V2 and MFCA fusion, the efficient fusion of frequency features and temporal dependencies in audio deepfake detection is achieved, providing new ideas and methods for deepfake detection.

\section{Related Work}

\subsection{Audio Deepfake Detection Technology	}

In recent years, deep learning technology has made significant progress in audio forgery detection\cite{9151013}. Through an end-to-end learning method , deep learning models such as CNN\cite{10574576},Resnet\cite{9265405}and so on, automatically extract audio features, thereby significantly improving the ability of audio deep forgery detection\cite{10434896}. However, existing deep learning models still mainly focus on the extraction of global features\cite{10136659}, and their ability to capture frequency features at fine granularity is still insufficient.

As an efficient deep learning model architecture, MobileNet has shown application potential in the field of audio forgery detection due to its lightweight and high-performance characteristics.\cite{8993089} In audio deepfake detection, MobileNet can be used to extract key features of audio signals that are crucial for identifying anomalies in forged audio. \cite{10704095}With the continuous advancement of technology and the continuous optimization of models, MobileNet is expected to play an increasingly important role in the field of audio deep forgery detection\cite{9859621}.

\subsection{Application of Mel-spectrogram in audio processing}
Mel-spectrogram, as a standard feature representation of audio signals\cite{10698112}, has been widely used in tasks such as speech recognition and audio classification. 
\begin{figure}[H]
	\centering
	\includegraphics[width=0.7\linewidth]{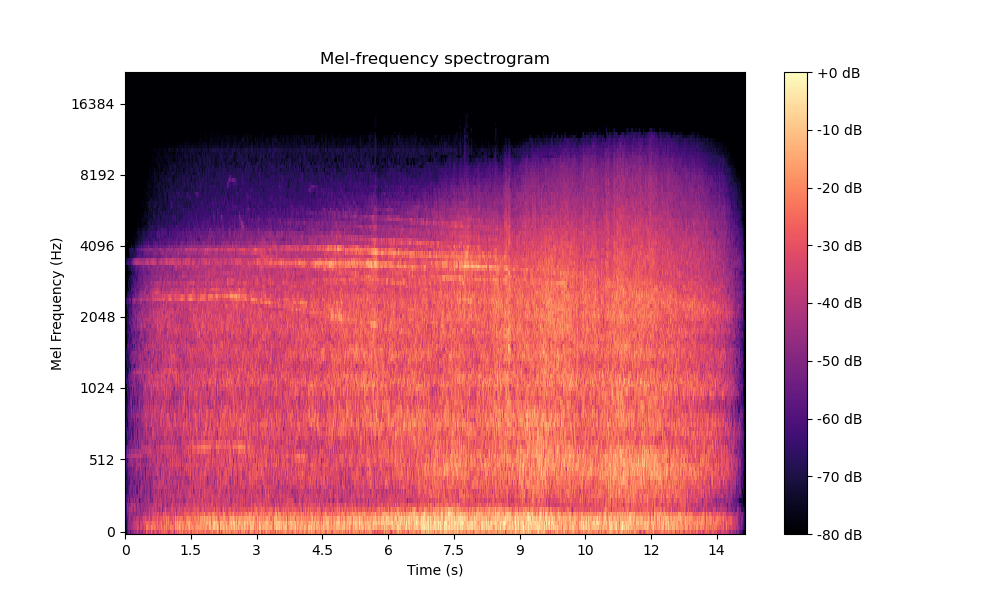}
	\caption{Deepfake audio mel spectrogram.}
	\label{mel1}
\end{figure}
Since Mel-spectrogram can effectively capture the frequency characteristics of audio signals\cite{10336152} and conform to the perception of human hearing, it has become the preferred input feature for audio deepfake detection. In recent years, researchers have tried to combine Mel-spectrogram with convolutional neural network (CNN) \cite{9613330}to further enhance its performance in audio deepfake detection. Some scholars\cite{8284159} have also conducted research by applying discrete cosine transform to the logarithm of the Mel filter signal to obtain the Mel frequency cepstral coefficient (MFCC) from the Mel spectrum.\vspace{-1.5em}

\begin{flalign}
	c(t, r)=\sum_{s=0}^{S-1} \log \left[X_{\mathrm{mel}}(t, s)\right] \cdot \cos \left[\frac{\pi \cdot r \cdot(s+0.5)}{S}\right] \quad \forall r=0, \ldots, R-1
\end{flalign}
where R is the number of DCT coefficients.

Despite this, existing research still has certain limitations\cite{10653050} in capturing multi-frequency domain information.

\subsection{Attention mechanism and its application in time series modeling}
Attention mechanisms have been widely used in the field of deep learning in recent years, especially the successful application of self-attention mechanisms\cite{Wang2021UCTransNetRT} and Transformer architectures in time series modeling. The self-attention mechanism can dynamically adjust the weights according to different parts of the input sequence, so as to better capture the relative importance of each part of the sequence\cite{8844925}. In audio deepfake detection, the attention mechanism is used to highlight the parts of the audio signal that are critical to the forgery features. In particular, the multi-frequency channel attention mechanism can automatically adjust its weights according to the characteristics of different frequency bands\cite{10656278}, thereby effectively enhancing the forgery recognition ability of the model.
In the context of multi-scale feature extraction\cite{sun2023multilevel}, the attention mechanism is combined with the multi-frequency conversion method to better capture the local and global contextual information in the audio.

\section{Method}

This study proposes an audio deepfake detection method based on the MFCA. This section will focus on the specific principles of this work.
\subsection{Data preprocessing}

In the audio processing process, we first need to extract the audio signal from the input wav format audio files. Since wav is a compressed format, the stored audio signals are losslessly encoded and cannot be directly input into the detection model for analysis. Therefore, we need to convert the wav format audio files into Mel spectrograms to capture the frequency domain features in the audio signal.

In order to calculate the spectrogram of the audio signal, we first need to split the original audio signal into multiple small frames and ensure that there is a certain overlap between adjacent frames.The processing of each audio frame can be expressed as:\vspace{-1em}

\begin{flalign}
x(t) \rightarrow\left\{x_{1}(t), x_{2}(t), \ldots, x_{M}(t)\right\}
\end{flalign}where each sub-segment $x_{i} (t)$ can be expressed as:\vspace{-1em}

\begin{flalign}
x_{i}(t)=x(t) \cdot \operatorname{rect}\left(\frac{t-t_{i-1}}{t_{i}-t_{i-1}}\right)
\end{flalign}where ${rect}(t)$ is the rectangular window function.

This frame segmentation method can effectively capture the changes in the signal over time, thus helping us analyze the time-varying characteristics of the audio signal. Through this process, the audio file is converted into a digitized audio signal for further processing.

\begin{figure}[H]
	\centering
	\includegraphics[width=0.6\linewidth]{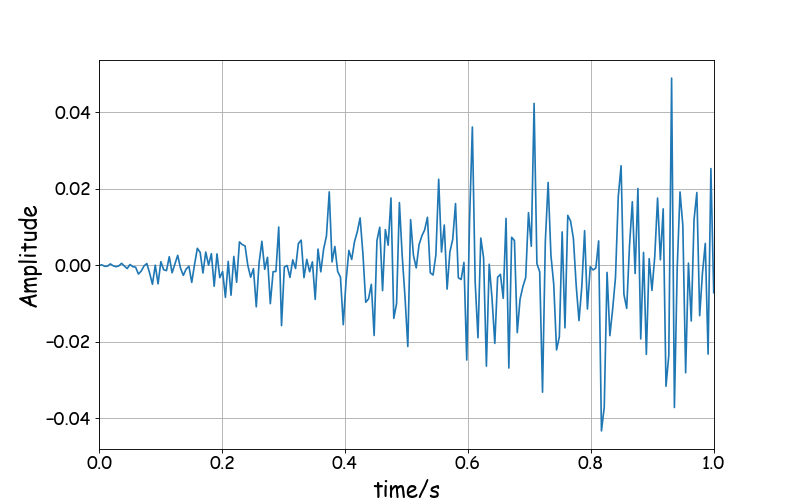}
	\caption{Image of the original signal extracted from deepfake audio}
	\label{fig:}
\end{figure}

Fourier transform is often used to convert signals from time domain to frequency domain, generally for continuous signals. \vspace{-1em} 
\begin{flalign}
	F(w) = \int f(t) e^{-jwt} \, dt
\end{flalign}where $f(t)$ is spectrum unknown signal,$F(w)$ is the spectrum function after Fourier transformation. Its independent variable is $w$ frequency,$e^{-jwt}$ is complex signal.

In the process of audio signal processing, since the sampling rate of the sampling device is limited, the sampled signals obtained are all discrete. Therefore, we use discrete Fourier transform (DFT) to convert the time domain signal into frequency domain representation.The sampling results $X[k]$ are as follows.\vspace{-1em}

\begin{flalign}
	X[k] = \sum_{n=0}^{N-1} x[t] e^{-j \frac{2\pi}{N} kt}
\end{flalign}where $N$ is the total number of sampling points, $j$ is imaginary unit, $k$ is the index in the frequency domain.

Then, we further process the spectrum. First, the frequency axis is mapped to a logarithmic scale to make it more consistent with the auditory perception characteristics of the human ear. At this point, the color dimension is mapped to the decibel (dB) scale, which enhances the visualization of the spectrogram in the low-frequency part.

After the above processing, we convert the audio signal into a Mel-spectrogram. This graphical spectrum representation can effectively capture the important frequency feature representations in the audio signal.
\begin{figure}[H]
	\centering
	\includegraphics[width=0.7\linewidth]{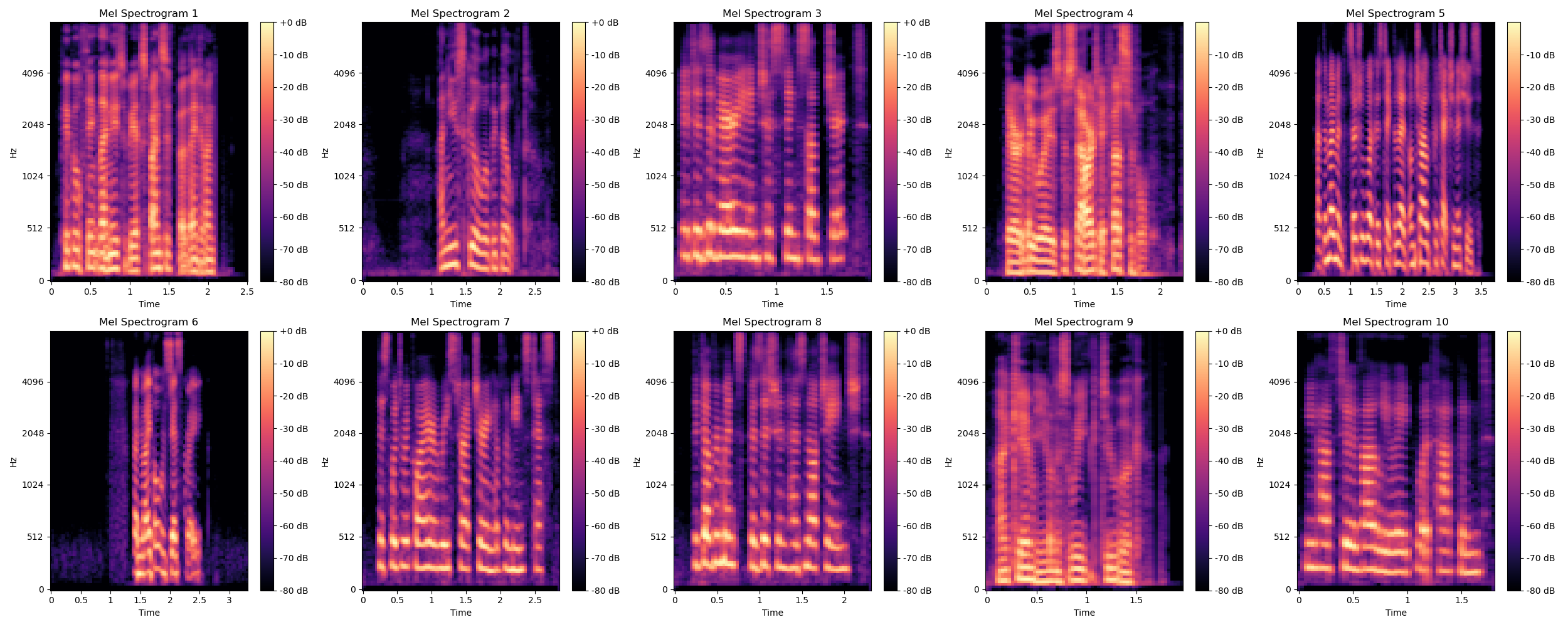}
	\caption{Mel spectrum converted from audio signal}
	\label{mel}
\end{figure}

\subsection{Feature extraction}
This project uses the MobileNet V2 model due to its superior performance in lightweight design and efficient feature extraction. MobileNet V2 was originally designed to provide an efficient, low-computation cost deep learning solution for resource-constrained devices (such as mobile devices), while combining powerful feature expression capabilities and high accuracy. Its main features include depth-separable convolution and inverted residual structure (Inverted Residuals), which not only significantly reduce model parameters and computational overhead, but also effectively improve the feature learning ability of the network. In addition, the inverted residual structure of MobileNet V2 optimizes the information flow by introducing a bottleneck layer, alleviates the vanishing gradient problem, and provides higher stability for feature extraction.

\textbf{MobileNet V2 Model.}

The core feature extraction of MobileNet V2 relies on deep separable convolution, which decomposes the traditional convolution into two steps:
first, channel-by-channel convolution (Depthwise Convolution), which performs convolution operations on each channel independently to extract local features and reduce redundant calculations; 
second, $1 \times 1$ convolution (Pointwise Convolution), which is responsible for fusing features from different channels to enhance the cross-channel feature interaction capabilities. This convolution method maintains efficient feature learning capabilities while greatly reducing the amount of calculation, and is particularly suitable for processing the time-frequency distribution of audio features.

\textbf{Feature enhancement of inverted residual structure}

During the feature extraction process, MobileNet V2 further optimizes feature expression with the help of inverted residual structure: The bottleneck layer expands the feature dimension through $1 \times 1$ convolution to enhance the nonlinear expression ability of the model. Residual connection retains important information in the input features through jump connections, optimizes information flow, and improves the stability of gradient propagation. In addition, the dimension compression operation reduces the feature dimension to a range suitable for downstream tasks, while reducing redundant calculations to ensure the full extraction and efficient processing of audio signal features.
\begin{figure}[H]
	\centering
	\includegraphics[width=0.7\linewidth]{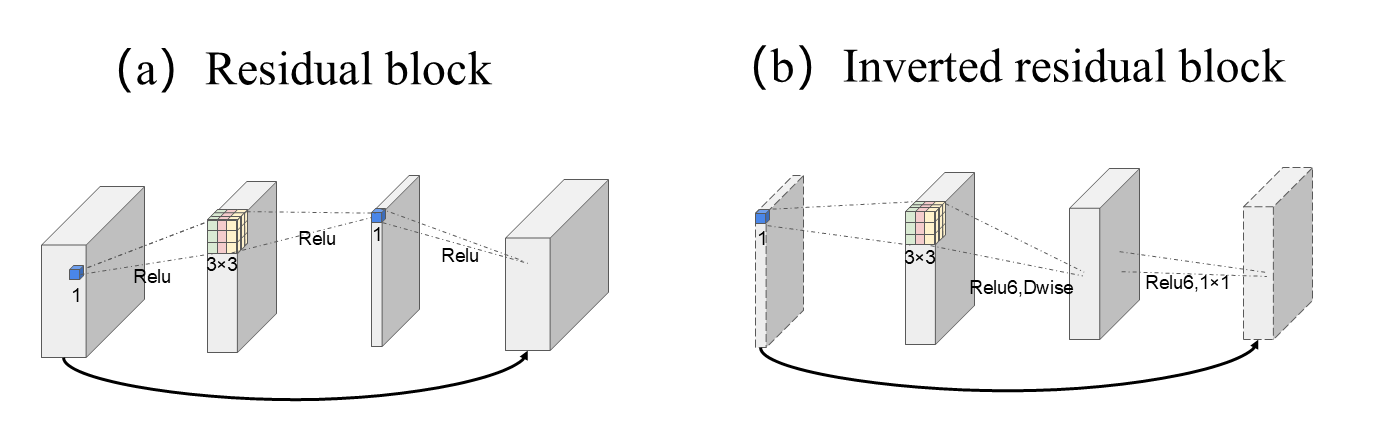}
	\caption{Comparison of residuals and inverted residuals}
	\label{mel}
\end{figure}
\textbf{Downsampling and multi-scale feature capture.} 

When processing spectrograms or MFCCs, MobileNet V2 achieves multi-scale feature capture through grouped convolution and multi-branch structures: low-frequency features are used to capture global information of the audio, such as rhythm and background characteristics; high-frequency features are used to extract detailed information, such as changes in pitch and short sound spikes. The model's downsampling operation gradually reduces the spatial resolution of the feature map, reduces computational complexity, and retains important information, thereby improving overall processing efficiency.

\textbf{Global Average Pooling}
In the final stage, MobileNet V2 uses the Global Average Pooling operation to compress the extracted two-dimensional feature map into a one-dimensional feature vector. This step effectively reduces the feature dimension while maintaining global semantic information, providing a concise and effective input for multimodal fusion, which helps improve the efficiency and accuracy of the model when processing multimodal data.

\subsection{Feature Fusion Based on MFCA and 2D DCT}

\begin{figure}[H]
	\centering
	\includegraphics[scale=0.28]{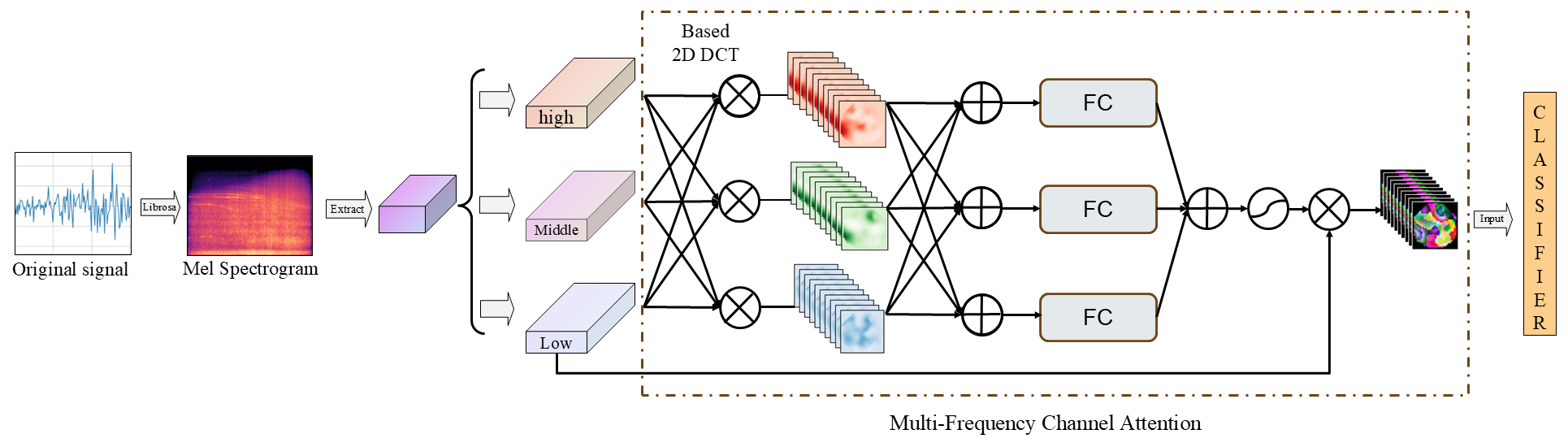} 
	\caption{Multi-frequency channel attention mechanism framework diagram}
	\label{mfca}
\end{figure}

After obtaining the deep features, the low frequency, medium frequency and high frequency in the features are first separated to obtain three different frequency bands. Each frequency band becomes a separate frequency channel. Then these low frequency, medium frequency and high frequency bands are used as inputs and passed into the MFCA module. MFCA assigns different weights to each frequency band and focuses on the frequency range that is most useful for audio deepfake detection. By separating the frequency bands, the model can focus on local features in different frequency ranges and can use the information of different frequency bands to make more accurate audio deepfake detection.

To further enhance the weight information of each frequency channel, the MFCA module introduces 2D DCT. DCT is an effective frequency domain analysis tool that can transform signals from time domain to frequency domain while keeping the main energy of the signal concentrated on a few frequency components.\cite{6781366} In MFCA, we apply 2D DCT to the attention map to extract statistics in the frequency domain. These statistics reflect the correlation between different frequency channels and their importance in the overall signal. By inversely transforming these statistics back to the time domain and combining them with the original attention map, we obtain an attention map enhanced in the frequency domain.

Finally, the MFCA module uses the enhanced attention map to weight the original Mel spectrogram to achieve feature fusion. Specifically, we multiply each element in the attention map with the Mel spectrogram feature at the corresponding position to emphasize important features and suppress minor features.

\section{Experiments}
In order to verify the effectiveness of the MFRN model proposed in this study, we conducted comparative experiments on multiple models. Experimental results show that with the introduction of MFCA, the performance of the MFRN model is significantly improved, especially showing stronger generalization ability when processing complex audio scenes.
\subsection{experiment setting}
\textbf{Dataset.} 
In this study, the dataset used to train our deepfake audio detection model is the "Fake or Real" dataset created by researchers at the University of York. The dataset consists of real and deepfake recordings and is released in four versions: for-original, for-norm, for-2sec, and for-rerec. The for-norm version was chosen for this study.

This dataset is balanced in terms of gender and category, and is standardized in terms of sampling rate, volume, and number of channels. These recordings have been used to train our model to effectively distinguish between real and fake samples. These samples cover a variety of deception methods, such as speech synthesis, voice conversion, and speech splicing, so that the generalization ability and robustness of the model can be fully evaluated.

\begin{table}[H]
	\caption{Fake or Real dataset distribution (file omitted)}
	\footnotesize
	\begin{center}
		\begin{tabular}{p{4cm}|p{2cm}}
			\hline
			Set  & Total \\ \hline                                                   
			Training    & 13956    \\
			Validation  & 2826\\
			Testing  & 1088\\
\hline 
		\end{tabular}
	\end{center}
	\label{tab300w}
	\vspace{-1em}
\end{table}

\textbf{Evaluation Metrics.} 
In order to comprehensively and accurately evaluate the performance of our proposed audio deepfake detection model combined with MFCA, we selected the following four key evaluation indicators: accuracy, precision, recall, and F1 score. These indicators can reflect the performance of the model from different perspectives and help us understand the advantages and disadvantages of the model more deeply.
\vspace{-1em}
\begin{flalign}
\textbf{Accuracy} 
\text{} = \frac{TP + TN}{TP + TN + FP + FN}
\end{flalign}
\vspace{-0.5em}\vspace{-1em}
\begin{flalign}
\textbf{Precision}
\text{} = \frac{TP}{TP + FP}
\end{flalign}
\vspace{-1em}\vspace{-0.3em}
\begin{flalign}
	\textbf{Recall}
	\text{} = \frac{TP}{TP + FN}
\end{flalign}
\vspace{-1em}\vspace{-0.3em}
\begin{flalign}
	\textbf{F1 Score}
	\text{} = \frac{\text{Precision} \cdot \text{Recall}}{\text{Precision} + \text{Recall}}
\end{flalign}

\textbf{Hyperparameter configuration.} 
The parameter configuration plays a crucial role in the design and optimization of neural networks. It not only precisely defines the structure of the network, but also significantly improves the network's performance by adjusting these configurations to meet specific application requirements. The following is a detailed explanation of the parameter configuration of neural networks:

\begin{table}[H]
	\caption{Hyperparameter configuration of neural networks}
	\footnotesize
	\begin{center}
		\begin{tabular}{p{4cm}|p{2cm}}
			\hline
			Parameter  & Parameter value  \\ \hline                
			batch\_size & 32 \\
			img\_height & 224 \\
			img\_width & 224 \\
			channels & 3 \\
			input\_shape & (224,224, 3) \\                        

			\hline 
		\end{tabular}
	\end{center}
	\label{tab300w}
	\vspace{-1em}
\end{table}

\subsection{Comparison with state-of-the-art methods}
To evaluate the performance of multi-frequency channel attention mechanism, we selected several existing benchmark models, such as CNN, VGG16, ResNet50, and MobileNet, for comparative experiments.CNN and VGG16 provide the key basis for audio deepfake detection with their powerful feature extraction capabilities\cite{10250463}, while ResNet50 extracts image features by stacking multiple residual blocks\cite{ren2024multimodalsentimentanalysisbased} to form a deep network to solve the deep network training problem, so that the model can learn audio features more deeply.These models have performed well in audio processing tasks \cite{9522625}and can automatically extract hierarchical representations of audio features\cite{8462062}, which are advantageous for processing audio data with temporal characteristics.The statistics of the results of the experiments are shown as follows.

\begin{table}[H]
\caption{ Comparisons on Fake or Real dataset. (\% omitted).}
\begin{center}
	\footnotesize
	\begin{tabular}{p{3.5cm}|p{1.2cm}p{1.2cm}p{1.2cm}p{1.4cm}p{1.4cm}p{1.2cm}p{1.1cm}}
		\hline
		Model  & Accuracy& Precision& Recall& F1 Score &epoch &batch\_size \\ \hline
		MobileNet		&81.1 &50.5	&49.7	&50.1  &10 &32  \\
		InceptionNet  &80.1 &50.2 &52.9 &51.5 &10 &32  \\
		VGG16		 &83.1 &\textbf{50.8} &42.5 &46.3 &10 &32  \\
		ResNet50		 &83.7 &50.4 &53.1 &51.7 &10 &32  \\		\hline
		\textbf{MFCMNet}	 &\textbf{88.2} &50.1 &\textbf{54.5} &\textbf{52.7} &10 &32  \\\hline
		CNN		 &79.3 &50.3 &50.6 &50.4 &20 &32  \\	
		\hline
	\end{tabular}
\end{center}
\label{tabFoR}
\end{table}

\subsection{ Experimental Results and Analysis.}
From the above experiments and comparisons, we draw the following conclusions and findings:

(1)In terms of accuracy, our MFCMNet(Multi-Frequency Channel MobileNet) model achieved significant results. This is a 4.5\% improvement over the baseline model, indicating that our MFCMNet model exhibits higher recognition capabilities and robustness when dealing with complex and diverse spoofing audio.

(2)Our model also performs well in terms of recall, which shows that our model is able to accurately identify the majority of deceptive samples while maintaining a low false positive rate.

(3)In addition, we also calculated the comprehensive evaluation index F1 score to further verify the performance of the model. The F1 score combines the information of accuracy and recall and can more comprehensively reflect the performance of the model in real application scenarios. Experimental results show that our MFCMNet model has also achieved significant improvements in F1 score, further verifying that our model has achieved a good balance between precision and recall, proving that this method is more effective in deceptive Effectiveness in speech detection tasks, with strong overall performance.

\section{Conclusion}

This study proposes an audio deepfake detection model MFCMNet based on the MFCA. This method significantly improves the accuracy and robustness of audio deepfake detection by combining the deep feature extraction capabilities of MobileNet V2 and the fine-grained feature capture capabilities of MFCA in the frequency domain. Experimental results show that the strategy of MFCA combined with MobileNet V2 performs best among all test methods and successfully achieves high-precision audio deepfake detection on the data set. . This innovative design not only achieves efficient integration of frequency characteristics and timing dependencies, but also provides new ideas and methods for audio deepfake detection.

The core contribution of this research is the innovative integration of MobileNet V2, MFCA and 2D DCT, and the successful application of this combination to the field of audio deepfake detection. With this approach, we significantly improve detection performance. In the future, we plan to further explore the integration of MFCA with other deep learning models in order to improve the adaptability and flexibility of the model while maintaining high performance.


\quad



\bibliography{mybibfile}

\end{document}